\documentstyle[12pt,epsfig,amssymb]{article}
\newcommand{\beq}{\begin{eqnarray}}
\newcommand{\eeq}{\end{eqnarray}}

\newcommand{\np}{Nucl.Phys.\ }
\newcommand{\pl}{Phys.Lett.\ }
\newcommand{\pr}{Phys.Rev.\ }
\newcommand{\asgen}{\alpha_s}

\newcommand{\as}{\alpha_{{}^{\widetilde{MOM}}}}
\newcommand{\asms}{\alpha_{{}^{\overline{MS}}}}
\newcommand{\MSB}{\overline{MS}}

\newcommand{\Lam}{\Lambda_{\widetilde{MOM}}}

\newcommand{\Gev}{{\rm GeV}}

\newcommand{\be}{\begin{equation}}
\newcommand{\ee}{\end{equation}}

\newcommand{\appsection}[1]{
\renewcommand{\thesection}{Appendix~\Alph{section}}
\section{#1}
\renewcommand{\thesection}{\Alph{section}}}

\def\np#1#2#3{Nucl.\ Phys.\ B#1 (19#3) #2}
\def\pl#1#2#3{Phys.\ Lett.\ #1B (19#3) #2}
\def\pr#1#2#3{Phys.\ Rev.\ D #1 (19#3) #2}

\def\prl#1#2#3{Phys.\ Rev.\ Lett.\ #1 (19#3) #2}

\def\zp#1#2#3{Zeit.\ Phys.\ C#1 (19#3) #2}
\begin{document}
\setcounter{page}{1}
\begin{flushright}
LTH 417\\
UPRF - 98 - 10\\ 
ULG-PNT-98-CP-1\\ 
\end{flushright}

\centerline{\bf{\huge Search for 
$1/p^2$ Corrections}}
\centerline{\huge   }
\centerline{\bf{\huge to the Running QCD Coupling}}  
\vskip 0.8cm
\centerline{\bf{ G. Burgio$^{a}$, F. Di Renzo$^{b}$,
C. Parrinello$^{b}$,
C. Pittori$^{c}$}} 
\centerline{$^{a}$ Dipartimento di Fisica, Universit\`a di Parma}
\centerline{and INFN, Gruppo Collegato di Parma, Parma, Italy}
\centerline{$^b$ Dept. of Mathematical Sciences, University of Liverpool}
\centerline{Liverpool L69 3BX, U.K.}
\centerline{(UKQCD Collaboration)}
\centerline{$^c$ Institut de Physique Nucl\'eaire Th\'eorique,} 
\centerline{Universit\'e de Li\`ege au Sart Tilmann} 
\centerline{B-4000 Li\`ege, Belgique.}
\begin{abstract}

We investigate the occurrence of power terms 
in the running  QCD coupling $\asgen(p)$ by analysing non-perturbative 
measurements at low momenta ($p \gtrsim  2 \Gev$) 
obtained from the lattice three-gluon vertex.
Our exploratory study provides some evidence for power contributions to 
$\asgen(p)$ proportional to 
$1/p^2$.
Possible implications for  
physical observables are discussed. 
\end{abstract}

\section{Introduction}
\label{sec:intro}
 
The standard procedure to parametrise non-perturbative QCD effects 
in terms of power corrections to perturbative results  
is based on the Operator Product Expansion (OPE). In this framework, 
the powers involved in the expansion are expected to be uniquely 
fixed by the symmetries and the dimension of the relevant operator 
product. It should be noted that,  
 due to the asymptotic nature of QCD perturbative 
expansions, power corrections are reshuffled between operators and 
coefficient functions in the OPE \cite{renormalons}.

The above picture has recently been challenged 
\cite{Akhoury,etc,Ceccobeppe}. It was pointed out that  
 power corrections which are not {\it a priori} expected from the OPE 
 may in fact appear in the expansion of physical observables. 
 Such terms may arise from (UV-subleading) power corrections to 
$\asgen(p)$, corresponding to non-analytical contributions to the 
$\beta$-function. To illustrate this point,
 consider for example a typical contribution to a 
condensate of dimension $2 \sigma$:

\begin{equation}
\label{intalfa}
\int_{\rho \Lambda^2}^{Q^2} \frac{dp^2}{p^2} (\frac{p^2}{Q^2})^\sigma 
\asgen(\frac{p^2}{\Lambda^2})\; .
\end{equation} 

A power contribution to $\asgen(\frac{p^2}{\Lambda^2})$ 
 of the form $(\frac{\Lambda^2}{p^2})^z$ would generate (from the UV 
limit of integration) a contribution to the condensate proportional to 
$(\frac{\Lambda^2}{Q^2})^z$. The fact that the dimension of such a term 
would be independent of $\sigma$ indicates that this contribution would be 
missed in a standard OPE analysis. 

Note that in the above manipulations $z$ could be 
in principle any (real) number. The value $z=1$ may in fact play a special 
role (see the discussion in Section \protect\ref{sec:relev}), 
as it would result in $\frac{\Lambda^2}{p^2}$ contributions to 
physical processes whose 
existence has been conjectured  for a long time, 
mainly in the framework of the UV renormalon \cite{trovare}. 

Clearly, the existence of $\sigma$-independent 
power corrections, if demonstrated, 
 would have a major impact on our 
understanding of non-perturbative QCD effects and may affect 
QCD predictions for several processes.
For example, $\frac{\Lambda^2}{Q^2}$ contributions 
may be relevant for the analysis of 
$\tau$ decays 
\cite{alt,Akhoury}. 

Although the size of such corrections could in 
principle be estimated directly from experimental data, 
it would be highly desirable to 
develop a theoretical framework where the occurrence of these effects 
is demonstrated and estimates are obtained from first principles QCD 
calculations. 
Some steps in this direction were performed in \cite{Lepage,Ceccobeppe}, 
where some evidence for an unexpected 
$\frac{\Lambda^2}{Q^2}$ contribution to the gluon condensate 
was obtained through lattice calculations. 

The aim of the present work is to test a method to detect  
the presence of power corrections in the running QCD coupling. 
Non-perturbative lattice estimates of the 
coupling at low momenta are compared with perturbative formulae. 
Although at this stage our work is exploratory in nature and further 
simulations will be required to obtain a conclusive answer, 
our analysis provides some preliminary evidence for power corrections. 
The final goal is to investigate the possible link between 
OPE-independent power corrections to physical observables and power 
terms in the running coupling.

%Our main aim in the present paper is to establish the 
%feasibility of our method. Indeed, as it will be discussed  
%in the following, since power corrections correspond to the interface between 
%perturbative and non-perturbative QCD, the lattice may provide a
%valuable computational tool.

The paper is organised as follows: in Section \protect\ref{sec:relev} 
we briefly review some theoretical arguments in support of power corrections 
to $\asgen (p)$, illustrating the special role that may be played by 
$\frac{\Lambda^2}{p^2}$ terms. 
In Section \protect\ref{sec:lattice} 
we explain the meaning of the lattice data and our strategy for the analysis. 
Some preliminary 
evidence for power corrections is discussed. 
Finally, in Section \protect\ref{sec:conc} we draw our conclusions. The 
appendix contains some technical details. 

\section{Clues for $\frac{\Lambda^2}{p^2}$ Corrections to $\asgen (p)$}
\protect\label{sec:relev}

Power corrections to $\asgen (p)$ can be shown to 
arise naturally in many physical schemes \cite{pino,maclep}.
The occurrence of such corrections 
cannot be excluded {\em a priori} in any renormalisation scheme. Clearly, 
given the non-analytic dependence of $(\frac{\Lambda^2}{p^2})^{z}$ terms 
on $\alpha_s$, power corrections cannot be generated or analysed in 
perturbation theory. In particular, 
the non-perturbative nature of such effects makes it very hard to 
assess their dependence on the renormalisation scheme, which is 
 only very weakly constrained by the general properties of the 
theory.

As discussed in the following,
despite the arbitrariness {\em a priori} of the exponent $z$, 
several arguments have been put forward in the past 
to suggest that 
a likely candidate for a power correction to $\asgen(p)$ would be a term 
of order ${\Lambda^2}/{p^2}$, i.e. $z=1$.

\subsection{Static Quark Potential and Confinement}
\label{sec:stat}
Consider the interaction of two heavy quarks in the static limit 
 (for a more detailed discussion see \cite{zakEQ}). 
In the one-gluon-exchange approximation, the static potential 
$V(r) $ can be written as
\begin{equation}
\protect\label{HQP}
V(r) \, \propto \, \alpha_s \ \int d^3k \,   
\frac{\exp^{i \vec{k} \cdot \vec{r}}}{|\vec{k}|^2}.
\end{equation}

Clearly the above formula yields the Coulomb potential $V(r) \approx 1/r$. 
Using standard arguments of renormalon analysis, one may consider a 
generalisation of (\protect\ref{HQP}) obtained by replacing $\asgen$ 
with a running coupling:
\begin{equation}
\protect\label{HQP2}
V(r) \, \propto \, \int d^3k \, \alpha_s (|\vec{k}|^2) \,  
\frac{\exp^{i \vec{k} \cdot \vec{r}}}{|\vec{k}|^2}.
\end{equation}

The presence of a power correction term of the form $\alpha_s(k^2) 
\propto {\Lambda^2}/{k^2}$ would generate a linear confining  
 potential $V(r) \sim K r$. 
Note that a standard renormalon analysis of (\protect\ref{HQP2}) 
(see \cite{zakEQ} for the details) reveals contributions to the 
potential containing various powers of $r$, but a linear contribution 
is missing. This is a typical result of renormalon analysis:  
renormalons can miss important pieces of non-perturbative information. 

\subsection{An Estimate from the Lattice}

The lattice community has been 
made aware for some time of 
the possibility of non-perturbative contributions to the running 
coupling; for a clear discussion see \cite{ChrisLAT94}. 
Consider the ``force" definition of the running coupling:
\begin{equation}
\alpha_{q\bar{q}}(Q) = \frac{3}{4} r^2 \frac{dV (r)}{dr} \;\;\;\;\;
(Q = \frac{1}{r}),
\end{equation}
where again $V(r)$ represents the static interquark potential.
By keeping into account the string tension contribution to $V(r)$, which 
can be measured in lattice simulations, one 
obtains a $1/Q^2$ contribution, whose order of magnitude is given 
by the string tension itself. Ironically, this term has been 
mainly considered as a sort of ambiguity, resulting in an 
indetermination in the value of $\alpha(Q)$ at a given scale. 
From a different point of view, such a term could be interpreted as a 
clue for the existence of a $\frac{\Lambda^2}{p^2}$ contribution, and it 
also provides an estimate for the expected order of magnitude of it, at 
least in one (physically sound) scheme.

\subsection{Landau pole and analyticity.}

It is well known that perturbative QCD formulae for the running of 
$\asgen$ inevitably contain singularities, which are often referred 
to as the Landau pole. The details of the analytical structure depend on 
the order at which the $\beta$-function is truncated and on the 
particular solution chosen. The existence of an interplay between the 
analytical structure of the perturbative solution and the structure of 
non-perturbative effects has been advocated for a long time 
\cite{RedBog}. To illustrate this idea, consider the 
one-loop formula for the running coupling  $\alpha_s (p)$:
\begin{equation}
\alpha_s(p^2)~=~
{1\over b_0 \ \log(\frac{p^2}{\Lambda^2})}.
\end{equation}
Here the singularity is a simple pole,  which can be removed if one 
redefines $\alpha_s (p)$ according to the following prescription:
\begin{equation}
\alpha_s(p^2)~=~
{1\over b_0 \ \log(\frac{p^2}{\Lambda^2})}+
{\Lambda^2 \over b_0 
(\Lambda^2-p^2)}, 
\end{equation}
where a power correction of the asymptotic form $\frac{\Lambda^2}{p^2}$ 
appears. However, the sign of such a correction is the opposite of 
what one would 
expect from 
the results of \cite{Ceccobeppe} and from the considerations in Section
\ref{sec:stat}, so that in the end one could envisage a more general 
formula for the regularised coupling: 
\begin{equation}
\alpha_s(p^2)~=~
{1\over b_0 \ \log(\frac{p^2}{\Lambda^2})}+
{\Lambda^2 \over b_0 
(\Lambda^2-p^2)}+c \ {\Lambda^2 \over p^2}. 
\label{eq:regu}
\end{equation}
The message from (\ref{eq:regu}) is that the perturbative 
coupling is {\em not defined} at the $\frac{\Lambda^2}{p^2}$ level, 
so the coefficient of the power correction is unconstrained, even after 
imposing the cancellation of the pole.

At higher perturbative orders one encounters multiple singularities, 
which include an unphysical cut. There are several ways to regularise 
them. In particular, the method discussed in \cite{RedBog} combines a 
spectral-representation approach with the Renormalization Group. The 
method was originally formulated for QED, but it has recently been 
extended to the QCD case \cite{Shirk}. 

Other approaches can be conceived to achieve a systematic regularisation 
of the singularities arising from the Landau pole, order by order in 
perturbation theory. In this way one obtains formulae for 
$\asgen(p)$  that are well-defined at all momentum scales. 

Such formulae would be quite useful in the framework of our study, since 
power corrections are expected to be sizeable at scales close to the 
location of the Landau pole. However, for the purpose of the preliminary 
investigation discussed in the present paper, we shall limit ourselves to
 a simpler approach, where one tries to fit the data by simply adding 
power corrections to the perturbative expressions, without attempting a 
regularisation of the Landau pole.

\section{Lattice Data and Power Corrections}
\protect\label{sec:lattice}

\subsection{$\asgen$ on the Lattice} 

Several methods for computing $\alpha_s (p)$ non-perturbatively on the lattice 
have been proposed in recent years \cite{Aida,Nara,Bali,Michael,io}. In 
most cases, the goal of such studies is to obtain an accurate prediction 
for $\asgen (M_Z)$, i.e. the running coupling at the $Z$ peak, 
which is a fundamental 
parameter in the standard model. For this reason, lattice parameters 
are usually tuned as to allow the computation of $\asgen (p)$ 
at 
momentum scales of at least a few \Gev s, where the two-loop asymptotic 
behaviour is expected to dominate and power contributions are suppressed. 
However, the same methods can in principle be applied to the study 
of $\asgen (p)$ at lower momentum scales, where power-like terms may be 
sizeable. For this purpose, the best method is one where 
one can  measure $\asgen (p)$ in a wide range of momenta from a
single Monte Carlo data set. 

One method which fulfills the above criterion is the determination of 
the coupling from the renormalised lattice three-gluon vertex 
function \protect\cite{io,cpcp}.  This is achieved by evaluating 
two- and three-point off-shell Green's functions of the gluon field 
in the Landau 
gauge, and imposing non-perturbative renormalisation conditions on them, for 
different values of the external momenta. 
By varying the renormalisation scale $p$, one can determine $\alpha_{s} (p)$ 
for different momenta from a single simulation. Obviously 
the renormalisation scale must be chosen in a range of lattice momenta 
such that both finite volume effects and discretisation  errors are under 
control. 
Such a definition of the coupling corresponds to a  
momentum-subtraction renormalisation scheme in continuum QCD \cite{???}. 
It should be noted that in this scheme the coupling is a gauge-dependent 
quantity. One consequence of this fact is that $1/p^2$ corrections should be 
expected, based on OPE considerations. We will return to this issue when 
drawing our conclusions. 
 
The numerical results for $\asgen(p)$ that we use for our 
investigation were obtained from 150 configurations 
on a $16^4$ lattice at $\beta=6.0$. 
 
For full details of the method we refer the reader to Ref. 
\protect\cite{cpcp}, where such results were first presented.
In order to detect violations of rotational invariance, different
combinations of lattice vectors have sometimes been used 
for a fixed value of $p^2$. This accounts for the graphical ``splitting" of 
some data points.

\subsection{Models for Power Corrections} 
As mentioned 
at the end of Section 2.3,
in the present work we shall not address 
the general problem  of 
defining a regular coupling at all scales. 
For the purpose of our preliminary investigation,
we shall compare the non-perturbative data for $\asgen$ 
with simple models obtained by adding a power correction term to the 
perturbative formula at a given order. 
In order to identify momentum intervals where our ansatz fits the data,  
one should keep in mind that the momentum range should start well above 
the location of the perturbative Landau pole, but it should nonetheless 
include low scales where power corrections may still be sizeable. 
The requirement of keeping the effects of the finite lattice 
spacing under control in the numerical data for $\asgen$ induces a 
natural UV cutoff on the momentum range. It is reassuring that intervals 
that fulfill these requirements can be identified, as specified in the 
following. 

One problem in this approach is the possible 
interplay between a description in terms of (non-perturbative) 
power corrections and our ignorance about higher orders of perturbation 
theory.
In particular, for the scheme that we consider, 
the three-loop coefficient of the $\beta$-function is not known. 
Knowledge of such a coefficient would allow to perform a more reliable 
comparison of our estimates for the $\Lambda$ parameter in our scheme 
with lattice determinations  of $\Lambda$ in a different scheme, 
for which the three-loop result is available \cite{Lusch}.  
In fact, although matching the $\Lambda$ parameter between different 
schemes only requires a one-loop computation (because of asymptotic freedom), 
the reliability of such a comparison rests on the assumption that the   
value of $\Lambda$ in each scheme is fairly stable with respect to the 
inclusion of higher orders, which in turn implies that a sufficient 
number of perturbative orders has been considered in the definition of 
$\Lambda$.  
In practice, when working at two- or three-loop order, the value of 
$\Lambda$ is still quite sensitive to the order of the calculation. 
For this reason, in the formulae for $\asgen(p)$ 
we shall append a subscript to the parameter $\Lambda$, to remind the 
reader that the value of such a parameter is expected to carry a sizeable 
dependence on the order of the perturbative calculation.
 
Even within such limitations, in the following we will argue that it is 
possible to estimate the impact of three-loop effects in our model and that 
a description with power corrections seems to be stable with 
respect to the inclusion of such effects.  

\subsection{Two-loop Analysis} 

At the two-loop level, we consider the following formula:
 
\begin{equation}
\protect\label{2lp}
\alpha_s(p) \, = \, \frac{1}{b_0 \, \log(p^2/\Lambda_{2l}^2)} \, - \, 
\frac{b_1}{b_0} 
\frac{\log(\log(p^2/\Lambda_{2l}^2))}{(b_0 \, \log(p^2/\Lambda_{2l}^2))^2}
\, + \, c_{2l} \, \frac{\Lambda_{2l}^2}{p^2}
\end{equation}

By fitting our data to (\protect\ref{2lp}) we obtain two sets of estimates 
for  the parameters ($\Lambda_{2l}$,$c_{2l}$), namely ($0.84(1)$,$0.31(3)$) 
and ($0.73(1)$,$0.99(7)$). The two results correspond to comparable 
values for $\chi^2_{dof}$, and in both cases we obtain $\chi^2_{dof} 
\leq 1.8$. 
%We notice that the second set of values  
%fits the data in a slightly larger momentum window, namely, 
%the point at $p \sim 1.8$ \Gev \ is best described 
%using the second set. 
In both cases, the momentum window extends up to $p \sim 3$ \Gev. 
We take the first set of values as our best estimate of the parameters 
as the corresponding value of $\Lambda_{2l}$ is close to 
what is obtained from a ``pure" two-loop fit, i.e. $\Lambda_{2l}$ is 
stable with respect to the introduction of power 
corrections.
Our choice for the value of $\Lambda_{2l}$ will be {\em a posteriori} 
supported also by independent considerations at the three-loop level. 
The momentum range that we are able to describe ($1.8 -3.0 \ \Gev$) is 
fully consistent with what one would expect from general 
considerations based on the value of the UV lattice cut-off 
and the value of $\Lambda_{2l}$. 
Notice that choosing between the two sets of values makes 
quite a difference in the UV region, where power effects are largely 
suppressed. 

In summary, a two-loop description with power corrections based on 
(\protect\ref{2lp}) 
fits well the data in a consistent momentum range. 
Our best fit of the data 
to (\ref{2lp}) is shown in Figure 1.
We were also able to check that if one tries to determine the exponent 
$z$ of the power correction $(\frac{\Lambda_{2l}^2}{p^2})^z$ from the 
fit, the best description of the data is obtained for $z \approx 1$.
We interpret this result as a confirmation of our theoretical prejudice 
$z=1$. However, one should note that since that the quality of our data 
makes a full three-parameter fit very hard, the above check 
of the value of $z$ and any other three-parameter fit 
that we mention in the following  sections were in fact obtained 
by performing a very large number of two-parameter fits, corresponding 
to different (fixed) values of the third parameter.

\begin{figure}[htb]\protect\label{fig:2loopot}
\begin{center}
\mbox{{\epsfig{figure=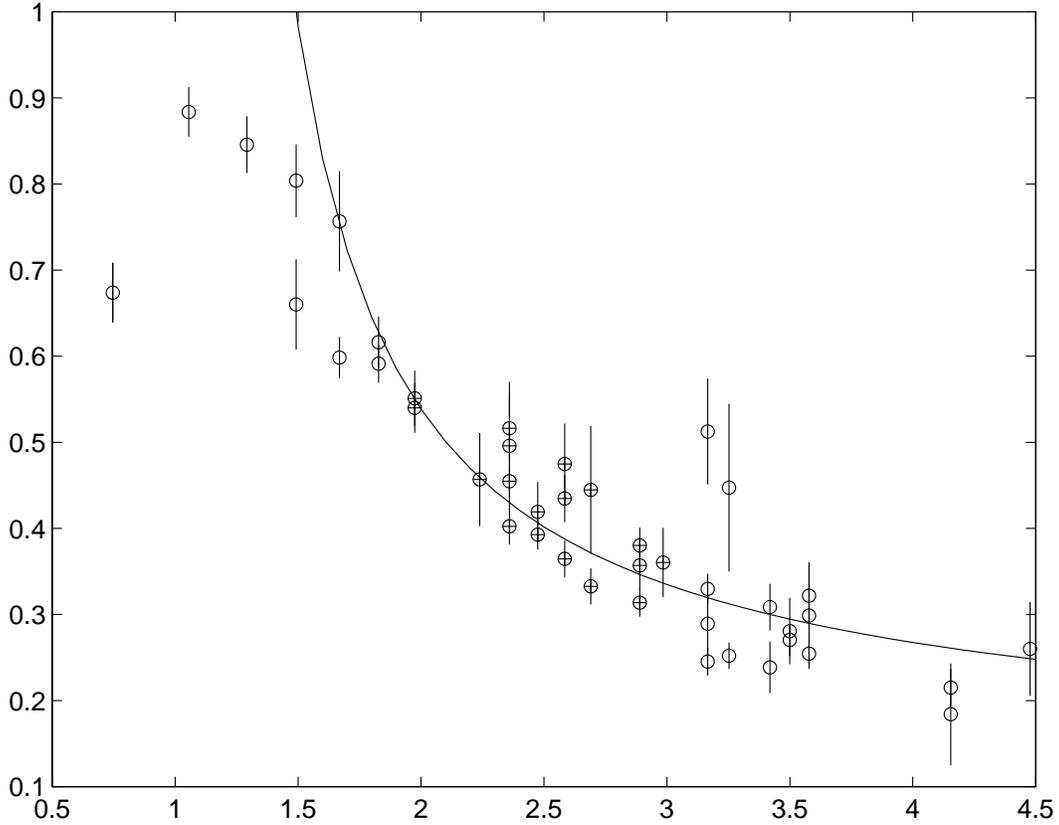,width=14.cm}}}
\end{center}
\caption{
The best fit to (\protect\ref{2lp}). 
The crossed-circled points indicate the fitting range.}
\end{figure}

\subsection{Three-loop Analysis} 
 
As already mentioned, a major obstacle for a three-loop analysisis is the 
fact that the first non-universal  coefficient $b_2$ of the perturbative 
$\beta$-function is not known for our scheme.

In order to gain insight, we start by performing a two-parameter fit to 
the standard three-loop expression for $\asgen(p)$, where the fitting 
parameters are $\Lambda_{3l}$ and the unknown coefficient $b_2$. We call 
$b_2^{eff}$ the fit estimate for $b_2$, to emphasise that we expect the 
effective value $b_2^{eff}$ to provide an order of magnitude estimate of 
the true (unknown) coefficient $b_2$.
Our best estimate for $\Lambda_{3l}$ and $b_2$ is $\Lambda_{3l} = 
0.72(1)$, $b_2^{eff} = 1.3(1)$, with $\chi^2_{dof} \approx 1.8$ 
(see the dashed curve in Fig. 2).
The error quoted for the fit parameters should always be interpreted 
within the effective description provided by the relevant formula.

The momentum range where we obtain the best description of the 
data is $p \sim 2 - 3$ \Gev. Our result for $\Lambda_{3l}$ provides 
(via perturbative matching) an estimate for $\Lambda_{\MSB}$, in 
very good agreement with the estimate in \cite{Lusch}, which was obtained
 from the computation of the $\Lambda$ parameter in a completely 
different scheme. 
Although both estimates are affected by our ignorance of higher loop 
effects, and our estimate also depends on the extra parameter 
$b_2^{eff}$, the agreement between the two results appears remarkable. 
In order to investigate the reliability of $b_2^{eff}$ as an estimate 
of $b_2$, we discuss in the appendix an argument which appears to provide a 
lower bound for the value of $b_2$ in our scheme, 
namely $b_2 \gtrsim 0.3$. 
Our value for $b_2^{eff}$ is therefore consistent with such a bound. 

Having obtained comparable values for $\chi^2_{dof}$ from 
the two-loop analysis with power corrections and from the ``pure" three-loop 
analysis, one may be led to consider our results as evidence against
the existence of power corrections, since so far they 
simply appear to provide an effective description of three-loop effects. 

However, we will argue now that there is room for power corrections even at 
the three-loop level. To this aim, consider the following three-loop 
formula with a power correction: 

\begin{eqnarray}
\protect\label{3lp}
\alpha_s(p) & = & \frac{1}{b_0 \, L} \, - \, 
\frac{b_1}{b_0} 
\frac{\log(L)}{(b_0 \, L)^2} \nonumber \\
& &  + \, \frac{1}{(b_0 \, L)^3} \, 
\left( \frac{b_2^{eff}}{b_0} + \frac{b_1^2}{b_0^2} 
(\log^2(L) - \log(L) + 1 ) \right) \nonumber \\
& &  + \, c_{3l} \, \frac{\Lambda_{3l}^2}{p^2},
\end{eqnarray}

where $L = \log(p^2/\Lambda_{3l}^2)$ and 
 $b_2^{eff}$ is again to be determined from a fit.

Fitting the data to (\protect\ref{3lp}), we obtain
$\Lambda_{3l} = 0.72(1)$, $b_2^{eff} = 1.0(1)$ 
and $c_3 = 0.41(2)$, with $\chi^2_{dof} \approx 1.8$, in a momentum range 
$1.8 - 3.0 \, \Gev$ (see Fig. 2). The above result 
was in practice obtained by performing a large number of two-parameter fits for 
$b_2^{eff}$ and $c_3$, for fixed values of $\Lambda_{3l}$. The range of 
trial values for $\Lambda_{3l}$ was suggested by the results of the ``pure" 
three-loop fit. 

We note the following:
\begin{enumerate}
\item 
the value for the scale parameter $\Lambda_{3l}$ 
is fully consistent with the previous determination from the ``pure" three-loop
description;
\item 
 the value for $b_2^{eff}$ is also reasonably 
stable with respect to the previous determination and it is also 
consistent with the approximate lower bound for $b_2$ discussed in the 
appendix;
\item 
by comparing results from fits to (\protect\ref{2lp}) and (\protect\ref{3lp}),  
it emerges that 
\begin{equation}
c_2 \Lambda_{2l}^2 = 0.22(2) \, \Gev^2 \sim c_3 \Lambda_{3l}^2 = 0.21(2)
 \, \Gev^2.
\protect\label{eq:tuttotiene}
\end{equation}
This approximate equality gives us confidence in the 
presence of power corrections, as it indicates that the 
 power terms providing the best fit to (\protect\ref{2lp}) and 
(\protect\ref{3lp}) are numerically equal.
In other words, there appears to be no interplay between
the indetermination connected to the perturbative terms and the 
power correction term, within the precision of our data, thus suggesting 
that a genuine $\frac{\Lambda^2}{p^2}$ 
correction is present in  the data. 
\end{enumerate}

\begin{figure}[hbt]\protect\label{fig:3loopot}
\begin{center}
\mbox{{\epsfig{figure=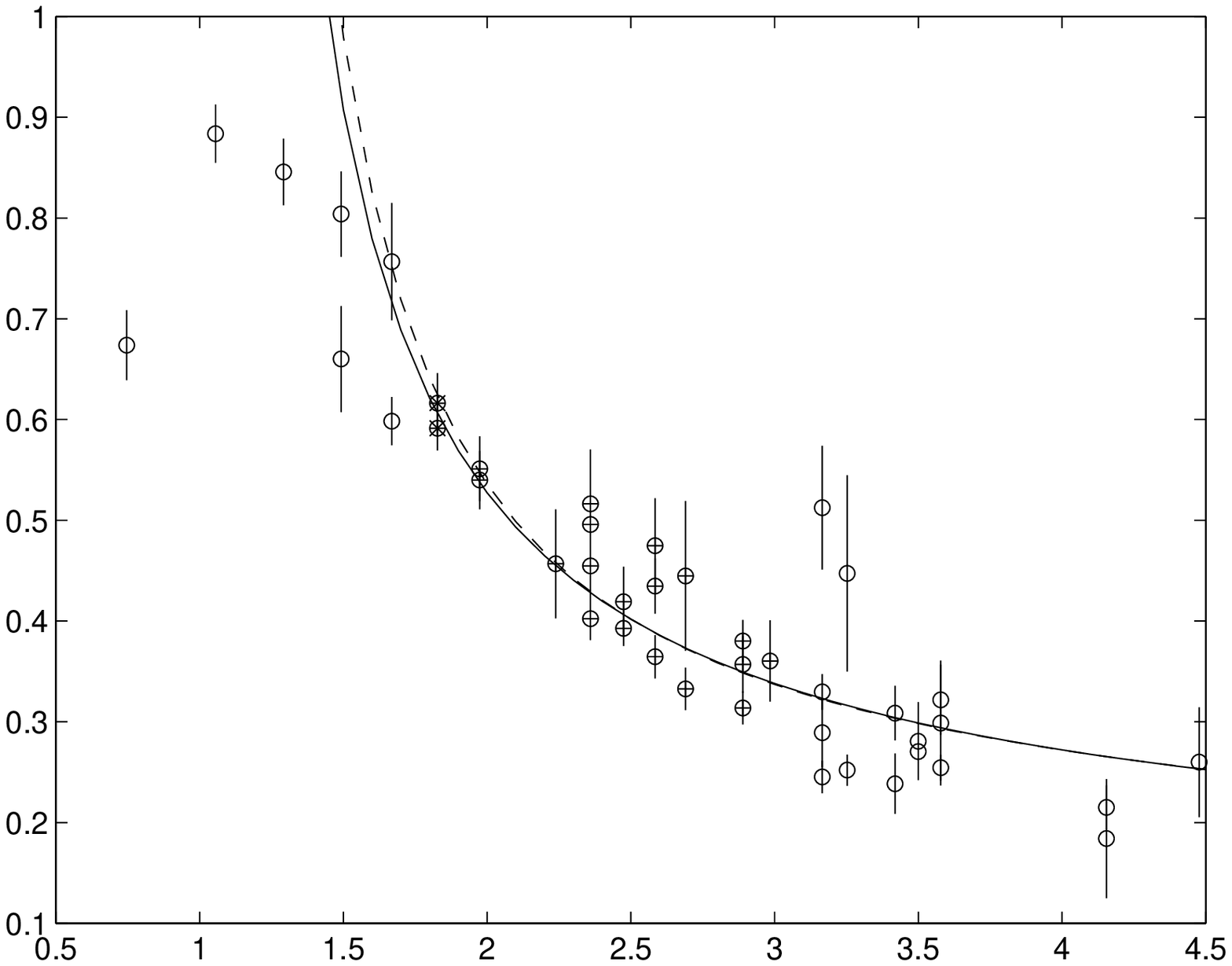,width=14.cm}}}
\end{center}
\caption{
Fits to (\protect\ref{3lp})
(solid line) versus a pure three-loop fit (dashed line). 
The crossed-circled points are consistent with both ansatze,
while the star-circled one is best fitted by (\protect\ref{3lp}).}
\end{figure}

Finally, the coefficient of the power correction is of the order of 
magnitude expected from the arguments in sections 2.1 and 2.2, that is,  
it is comparable to the standard estimate for the string tension squared. 

One may argue at this point that at the two-loop level we had to choose 
between two sets of values for ($\Lambda_{2l}$,$c_{2l}$), and that our 
choice is crucial for the validity of (\protect\ref{eq:tuttotiene}). 
 An {\it a posteriori} justification for our choice can be obtained from the 
following test:
we plot a few values for $\asgen(p)$ as 
generated by the ``pure'' three-loop formula 
for $\Lambda_{3l}= 0.72$ and $b_2 = 1.0$. Then, by fitting such 
points to the ``pure'' two-loop formula, one gets $\Lambda_{2l} 
\approx 0.84$, i.e. the value for which (\protect\ref{eq:tuttotiene}) holds.

%The above test seems to confirm our picture that perturbative and 
%non-perturbative (power) contributions do not mix in our formulae  
%when upgrading from a two-loop to a three-loop 
%description, 
%\footnote{We have checked that a four-loop analysis is not 
%viable given the quality of the numerical data.}, 

\section{Conclusions}
\protect\label{sec:conc}

We have discussed an exploratory investigation of power corrections 
in the running QCD coupling $\asgen(p)$ by comparing non-perturbative 
lattice results with theoretical models. Some evidence was 
found for $1/p^2$ corrections, whose size was consistent  
with what is suggested by simple arguments from the static potential.

At the technical level, our results need further confirmation from the 
analysis of a larger data set and a study of the dependence of the fit 
parameters on the ultraviolet and infrared lattice cutoff. 
%However, assuming our results are confirmed, 
%they would provide strong evidence in support of 
%the conjecture that power corrections to $\asgen(p)$ may be responsible for  
% the appearance of ``anomalous" (i.e., not accounted for by OPE) 
%power corrections into current correlation functions and physical observables.
Assuming our findings are confirmed at the technical level, 
one needs to address the issue of assessing the 
scheme dependendence of our results. As already discussed, the 
non-perturbative nature of power corrections makes it very hard to 
formulate any theoretical procedure to estimate the impact of scheme 
dependence. 
The best one can do at this stage is to consider 
different renormalisation schemes and definitions of the coupling and 
gather numerical evidence and 
formal arguments supporting power corrections to $\asgen(p)$.
In this way, scheme-independent features may eventually be identified. 
For example, on the basis of our results, we note the following:
\begin{itemize}
\item Theoretical arguments suggest $1/p^2$ corrections both for 
the coupling as defined from the static potential and for the one 
obtained from the three-gluon vertex. The arguments for the former case 
were outlined in Sections 2.1 and 2.2. As far as the coupling from the 
three-gluon vertex is concerned, $1/p^2$ corrections appear in an 
OPE analysis if one keeps into account the fact that such a coupling is 
{\it a priori} gauge dependent, so that a dimension 2 condensate appears 
in the relevant OPE. 

\item In the static potential case, the theoretical argument also 
provides an estimate for the order of magnitude of the coefficient 
of the $1/p^2$ correction, while in the three-gluon vertex case the 
OPE argument provides no estimate for it, suggesting instead that it 
may depend on the gauge.  
However, our numerical result in the Landau gauge is in striking 
agreement with the estimate for the static potential case. Although 
such an agreement may of course be accidental, it calls for further 
investigation, which may be performed by attempting a similar 
calculation in a different gauge. 
\end{itemize}

The issue of scheme dependence will 
be the focus of our future work.

\section{Acknowledgements} 
We thank B. Alles, H. Panagopoulos and D. G. Richards for allowing us 
to use data files containing the results of Ref. \protect\cite{cpcp}.   
C. Parrinello acknowledges the support of PPARC
through an Advanced Fellowship.
C. Pittori thanks J. Cugnon and the ``Institut de 
Physique de l'Universit\'e de Li\`ege au Sart Tilman" 
and acknowledges the partial support of IISN.
We thank C. Michael for stimulating discussions. 

\setcounter{section}{0}
\appsection{}

Consider the perturbative matching between our scheme and the $\MSB$ 
scheme 

$$ \as \, = \, \asms \, + \, c_1 \, \asms^2 \, + \, c_2(b_2) \, 
\asms^3 
\, + \, O(\asgen^4) $$

As it is well known, $c_1$ determines the ratio of the $\Lambda$ 
parameters in the different schemes, while $c_2$ depends on $c_1$ and the 
difference between the value of $b_2$ in our scheme and $b_2^{\MSB}$. 
We assume that at very high momentum values 
 ($p > 150$ \Gev) the running coupling follows the three-loop 
asymptotic formula.
Then if one takes the value for $\Lambda_{\MSB}$ from \cite{Lusch} and 
 the value for 
$\Lam$ in our scheme from the perturbative matching,
the only unknown parameter in the above expression 
 is the value of $b_2$ in our scheme.
By demanding that at the two-loop level the expansion of one coupling in 
powers of the other is still convergent (i.e. the convergence is better at 
two loops than at one loop as the series are not yet displaying their 
asymptotic 
nature) we obtain an approximate lower bound for the unknown 
coefficient as $b_2 \gtrsim 0.3$. We have checked that such a technique 
provides sensible results for every couple of couplings for which a two-loop 
matching is known.

\end{document}